\documentclass[aip,pra,amsmath,amssymb,reprint,superscriptaddress,floatfix]{revtex4-1}

\pdfoutput=1
\usepackage{graphicx}  
\usepackage{bm}        
\usepackage{amssymb}   
\usepackage{lettrine}
\usepackage{soul}
\usepackage{color}
\usepackage{braket}
\usepackage{tabularx}
\usepackage{subcaption}

\begin{document}

\title{Precision Measurement of a low-loss Cylindrical Dumbbell-Shaped Sapphire Mechanical Oscillator using Radiation Pressure}
	
\author{J. Bourhill}
\email{jeremy.bourhill@uwa.edu.au}
\author{E. Ivanov}
\author{M.E. Tobar}	

\affiliation{ARC Centre of Excellence for Engineered Quantum Systems, University of Western Australia, 35 Stirling Highway, Crawley WA 6009, Australia}

\date{\today}


\begin{abstract}
\noindent \textit{We present first results from a number of experiments conducted on a 0.53 kg cylindrical dumbbell-shaped sapphire crystal. This is the first reported optomechanical experiment of this nature utilising a novel modification to the typical cylindrical architecture. Mechanical motion of the crystal structure alters the dimensions of the crystal, and the induced strain changes the permittivity. These two effects result in parametric frequency modulation of resonant microwave whispering gallery modes that are simultaneously excited within the crystal. A novel low-noise microwave readout system is implemented allowing extremely low noise measurements of this frequency modulation near our modes of interest, having a phase noise floor of -165 dBc/Hz at 100 kHz. Fine-tuning of the crystal{'}s suspension has allowed for the optimisation of mechanical quality factors in preparation for cryogenic experiments, with a value of \textit{$Q=8 \times 10^7$} achieved at 127 kHz. This results in a $Q\times f$ product of \textit{$10^{13}$}, equivalent to the best measured values in a macroscopic sapphire mechanical system. Results are presented that demonstrate the excitation of mechanical modes via radiation pressure force, allowing an experimental method of determining the transducer{'}s displacement sensitivity $df/dx$, and calibrating the system. Finally, we demonstrate parametric back-action phenomenon within the system. These are all important steps towards the overall goal of the experiment; to cool a macroscopic device to the quantum ground state at millikelvin temperatures.}
\end{abstract}

\maketitle

\section{Introduction}
\lettrine[nindent=0em,lines=3]{T} \\
\\he field of optomechanics is producing many exciting results, from extremely precise sensors with applications in a wide variety of fields, including sensitive detection of previously immeasurable signals \cite{fields1,fields2,fields3,fields4,fields5,fields6,fields7}, quantum information processing \cite{fields8,fields9},and tests of fundamental quantum theories, including potential tests of quantum gravity \cite{quantgrav1,LIGO}. The majority of work in this area is based around microscopic or mesoscopic resonators, which have the advantage of extremely low mass and therefore relatively large optomechanical coupling factors, allowing for a plethora of interesting physics to be investigated. In particular, the quantum regime of mechanical motion (the so-called {``}standard quantum limit{''}) is far more accessible for these extremely small resonators. Great challenges arise when dealing with macroscopic resonators, and it has been shown that state-of-the-art technologies are required to approach observations of the quantum world in such resonators, due to their significantly larger masses \cite{clayton1,claytonthesis}. \\
\indent The essential requirement for the observation of optomechanical effects is to generate coupling between phonons and photons within a resonant structure. This coupling can take many forms, from variations in path length of a Fabry-Perot resonator via a moveable mirror (such as the LIGO experiment) \cite{LIGO,fabry-perot}, oscillations of a cantilever capacitively coupled to two electrodes \cite{cantilever}, ``membrane-in-the-middle" type experiments \cite{membrane}, or as in the present work, crystalline resonators, which can support both optical and microwave resonant fields and mechanical modes \cite{crystal1,crystal2}. All of these systems function via mechanical effects causing a disturbance to an optical field, which can be measured as a frequency shift. In the case of a macroscopic whispering gallery mode (WGM) resonator, its frequency is sensitive to changes in path length (i.e. the circumference and height of the cylinder), and strain in the crystal lattice, both of which are induced by mechanical motion. The latter effect is in fact the more dominant mechanism\cite{clayton1,claytonthesis}. These lattice vibrations can be caused by thermal phonons within the resonator at room temperature.

\section{Sapphire Resonator: Geometry and Modes of Vibration}
\begin{figure}[h!]
\captionsetup[subfigure]{justification=centering}
    \centering
      \begin{subfigure}{0.3\textwidth}
        \includegraphics[width=\textwidth]{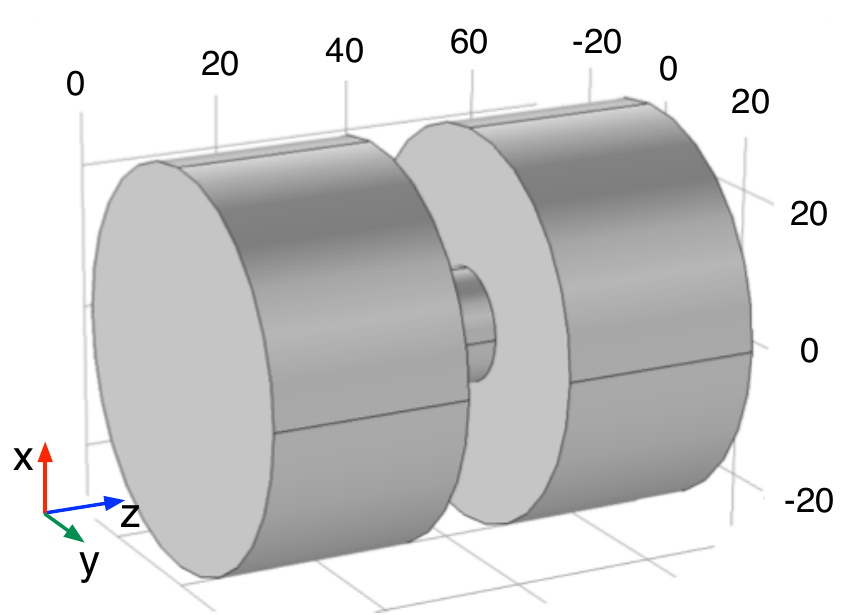}
          \caption{}
          \label{fig:bar}
      \end{subfigure}
     \begin{tabular}{c}
	      \begin{subfigure}{0.15\textwidth}
	        \includegraphics[width=\textwidth]{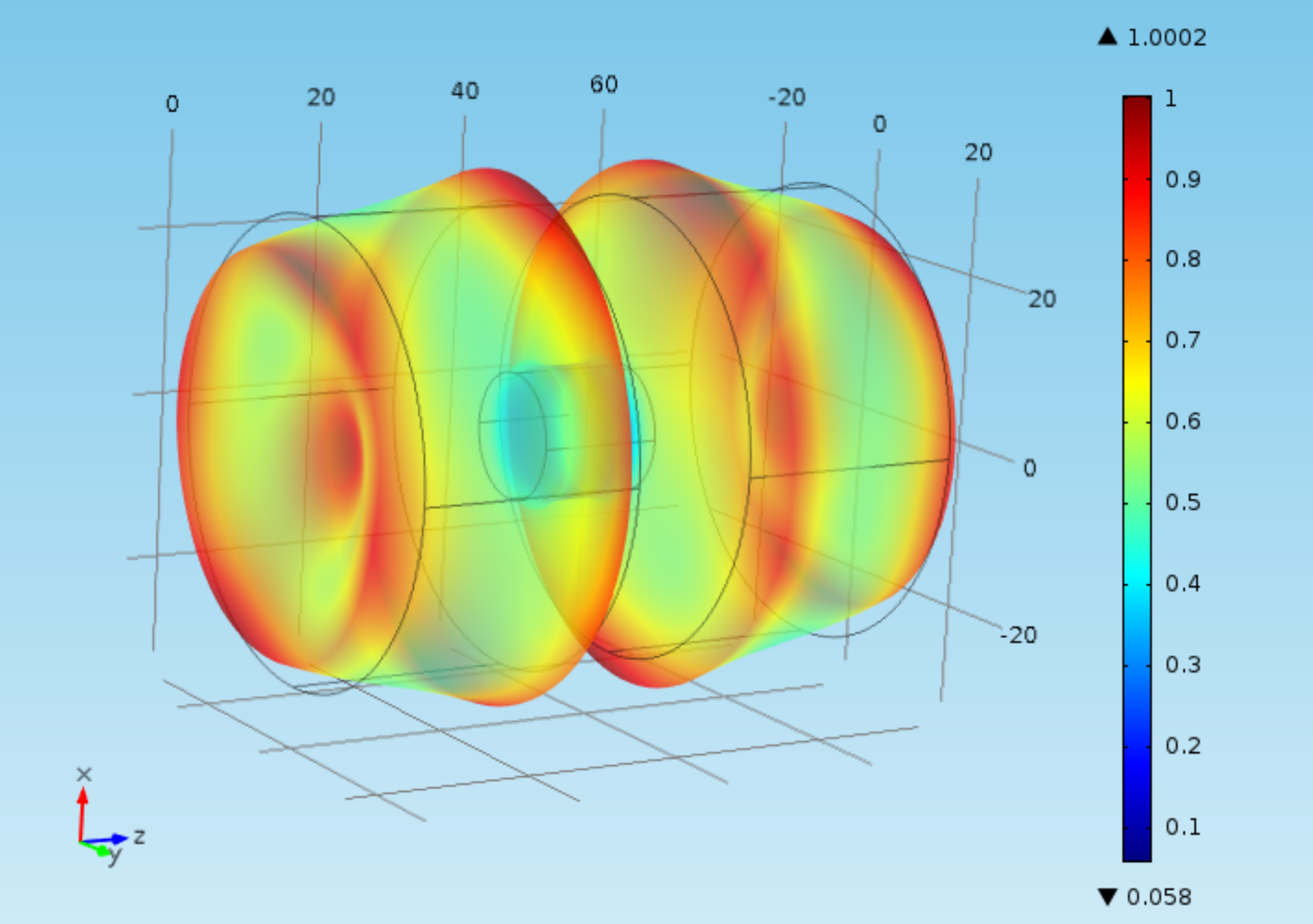}
	          \caption{}
	          \label{fig:95}
	      \end{subfigure}\\
	      \begin{subfigure}{0.15\textwidth}
	        \includegraphics[width=\textwidth]{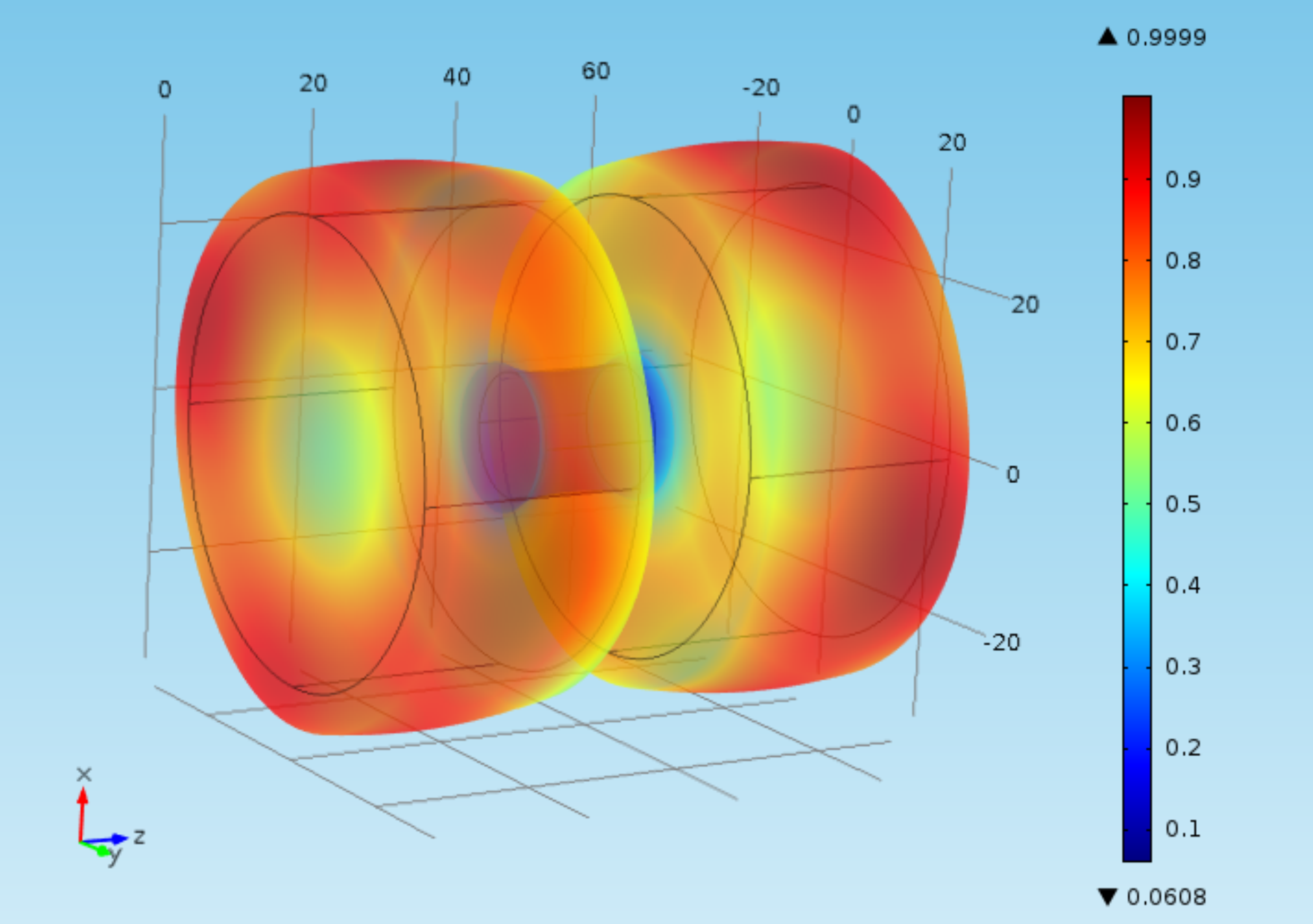}
	          \caption{}
	          \label{fig:127}
	      \end{subfigure}
	\end{tabular}
	\caption{(a) Graphical representation of the SB resonator modelled using FEM. Lengths shown are in millimetres. Normalised displacement field of the 95 kHz (b) and 127 kHz (c) resonant mechanical modes.}
\end{figure}
\noindent The sapphire resonator studied takes the form of a dumbbell-shaped cylinder and will herein be referred to as the {``}Split Bar{''} (SB) resonator, with its dimensions outlined in Fig. \ref{fig:bar}. It is suspended via a wire loop around the central {``}neck{''}. This shape was chosen to isolate the point of suspension from the mechanical motion in order to maximise mechanical Q-factors. The crystal is high quality HEMEX grade sapphire and was grown using the heat exchange method by Crystal Systems, USA, cut to dimensions and then optically polished.\\
\indent The SB is both an electromagnetic and mechanical resonator. Its cylindrical shape permits high Q-factor WGMs to be excited at microwave frequencies, and via the mechanism explained above the coupling between acoustic and electromagnetic modes required for the observation of optomechanical effects is achieved. \\


\noindent The SB resonator was modelled using Finite Element Modelling (FEM) software (COMSOL\texttrademark) to identify suitable mechanical modes within the crystal. Given the non-uniformity of the dimensions, the eigenfrequencies of the resonant modes were solved for varying neck diameters, ranging from the limiting cases of two separate cylinders of height 27.4 mm, to a single uniform cylinder of height 63.75 mm, in order to identify the correct families of the modes. 

\begin{figure}[h!]
	\centering
		\includegraphics[width=0.45\textwidth]{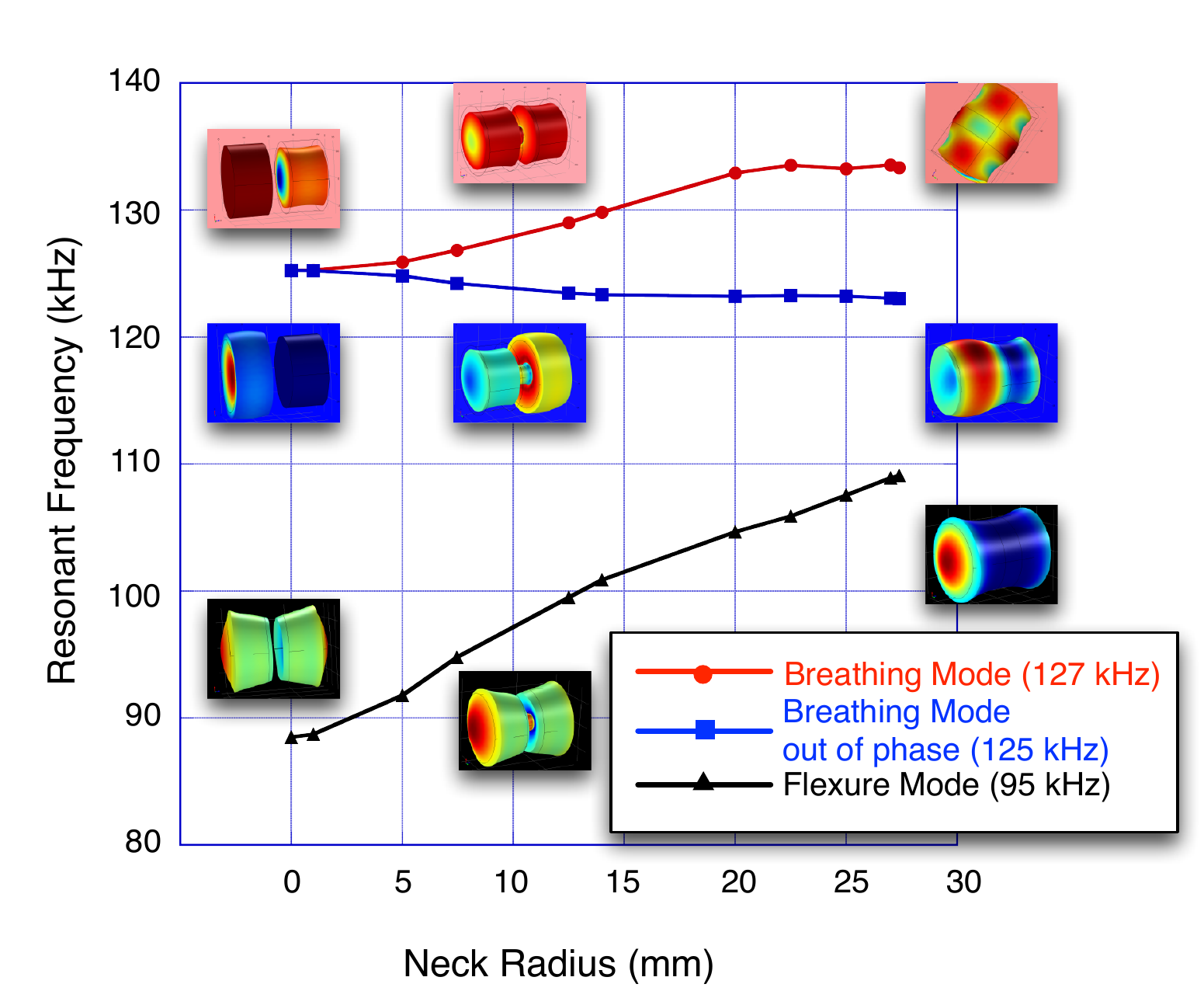}
	\centering	
		\caption{Three relevant mechanical mode families resonant frequencies as a function of neck diameter. The SB resonator has a neck diameter of 14.92 mm. By examining the two limiting cases of the SB (i.e neck diameter $= 0$ mm; corresponding to two separate cylinders, and neck diameter $=$ 27.4 mm; corresponding to one large cylinder, the origin of each mode can be correctly identified.)}
		\label{fig:mechanical}
\end{figure}

\noindent From Fig. \ref{fig:mechanical}, we can see that the in-phase and out of phase breathing modes (127 kHz and 125 kHz modes, respectively) are degenerate for the case of two separate crystals, but when they become joined, their frequencies tune in opposite directions. These modes become very different in the limiting case of a full cylinder. Each one of the modes in Fig. \ref{fig:mechanical} has been detected with frequencies in very good agreement with the modelled values ($<1\%$). However, this article will mainly deal with the 95 kHz mode (black curve) and the 127 kHz mode (red curve), as they have the two highest Q-factors. The low Q-factor observed for the 125 kHz mode is predicted by the modelling due to a large amount of displacement at the suspension point on the neck of the dumbbell, resulting in larger suspension losses. 

From the FEM, it is possible to obtain deformation gradients ($\frac{dz}{z}$, $\frac{dr}{r}$, $\frac{d\phi}{\phi}$) and total displacement curves for each of the modes, allowing an estimate of the WGM frequency sensitivity to each different mechanical mode family. For simplicity, this is done using a model of only one cylindrical end of the SB, at the mode of interest{'}s corresponding eigenfrequency in this limiting case, as depicted in Fig. \ref{fig:mechanical}.\\


\noindent To determine the magnitude of frequency variations due to mechanical vibrations ($\frac{df}{dx}$) of any particular WGM, we expand on the method used by Locke \textit{et al.} \cite{clayton1}. The frequency of WGMs in cylindrical sapphire, assuming the crystal axis (c-axis) is aligned with the z-axis, is dependent upon four variables; the permittivities of the crystal perpendicular and parallel to its c-axis, $\varepsilon_\perp$ and $\varepsilon_\parallel$, respectively (sapphire is an anisotropic crystal), and the crystal's dimensions (diameter, $D$ and length, $L$). \\
\indent The frequency sensitivity of the WGM can be calculated from

\begin{multline}
\displaystyle \frac{z}{f} \frac{d f}{d z}= \nu\left(\left(M_r p_{\varepsilon_r} +M_\phi p_{\varepsilon_\phi}\right)K_{\varepsilon_\perp}+p_D\right)\\
-M_z p_{\varepsilon_z} K_{\varepsilon_\parallel}-p_L,
\label{eq:dfdx}
\end{multline}

\noindent where $p_i$ are the normalised tuning coefficients of the WGM frequency with respect to the variable $i$ (i.e. $p_i=\left |\frac{\delta f^{res}}{\delta i}\right |\frac{i}{f}$), $\nu$ is Poisson{'}s ratio, $K_{\varepsilon_i}$ represents the strain dependence of permittivity (i.e. $K_{\varepsilon_\parallel}=\frac{d \varepsilon_\parallel}{d z}\frac{L}{\varepsilon_\parallel}$ and $K_{\varepsilon_\perp}=\frac{d \varepsilon_\perp}{d r}\frac{D}{2\varepsilon_\perp}$), and $M_i$ is the displacement modification factor, and represents the overlap of electric field ($\underline{E_i}(r,z,\phi)$) and strain ($S_i(r,z,\phi)$) in the $i$ direction; 

\begin{equation}
M_i=\frac{\int_V S_i(r,z,\phi) \varepsilon_i \underline{E_i}(r,z,\phi)\underline{E_i^*}(r,z,\phi) dr dz d\phi}{N_i},
\end{equation}
where $N_i$ is a normalisation constant. \\

\noindent Frequency sensitivity to changes in permittivity only occurs in the electrical component of the resonant microwave mode. This is due to the dependence of the E-field on $\varepsilon$ compared to a B-field{'}s dependence on $\mu$, a direct result of Maxwell's equations. WGMs have two main polarisations - WGH modes with dominant $E_z$, $H_r$ and $H_\phi$ field components, and WGE modes with dominant $H_z$, $E_r$ and $E_\phi$ components. Therefore, transduction in the split bar will be primarily due to the strain induced in the direction of the WGM{'}s E-fields. Since $\nu =0.3$ for sapphire, WGH modes are more strongly affected by mechanical motion.\\

\begin{figure}[h!]
\centering
	\includegraphics[width=0.45\textwidth]{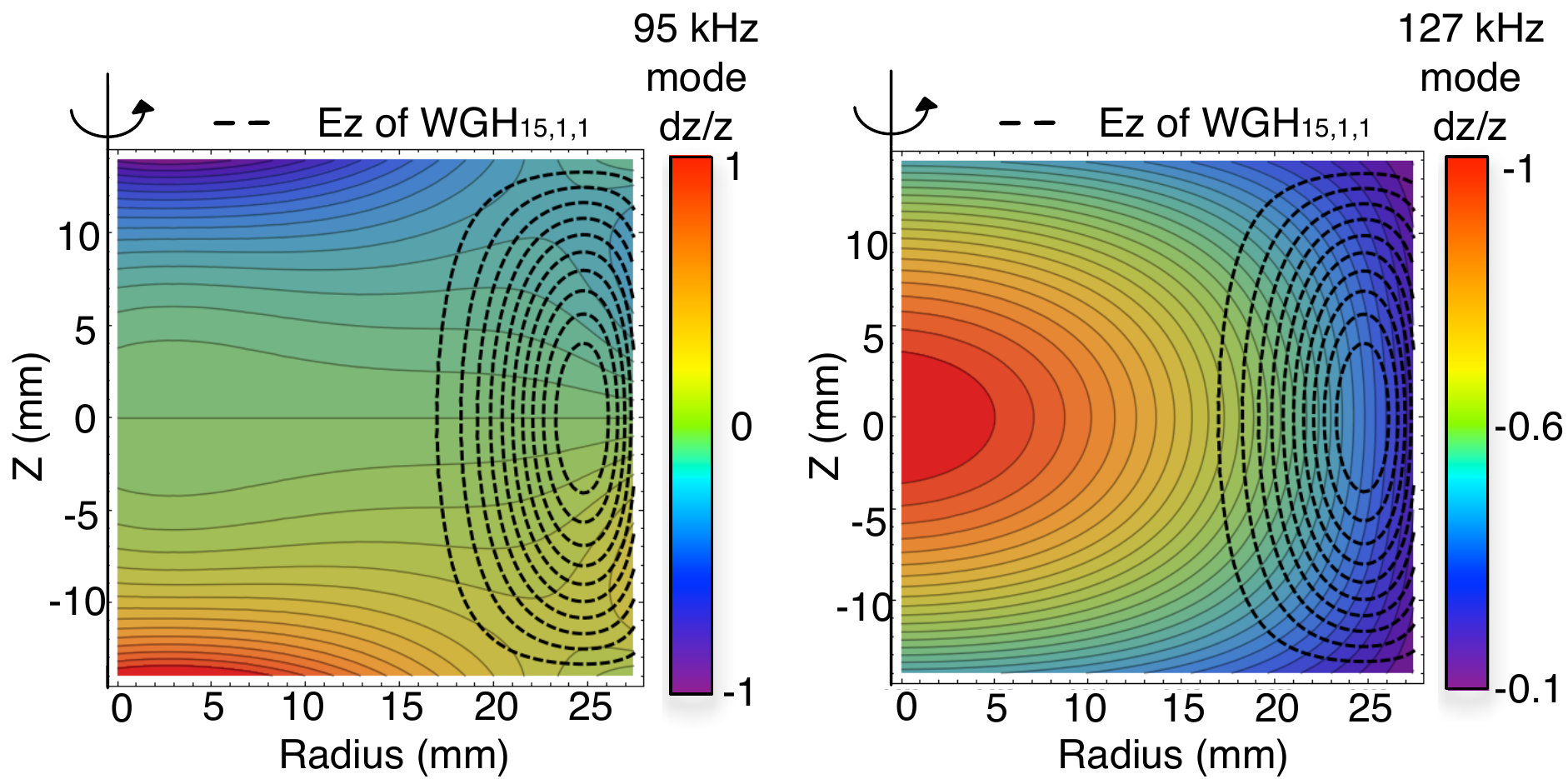}
	\caption{Overlap of WGH$_{15,1,1}$ $E_z$ field and deformation gradient $\frac{dz}{z}$ used to calculate $M_z$ modification factors in equation (\ref{eq:dfdx}).}
	\label{fig:Mz}
\end{figure}

\noindent Figure \ref{fig:Mz} depicts the method by which the $M_z$ factors are calculated for the 95 kHz and 127 kHz mechanical modes. The overlap of the electromagnetic and strain fields determines the magnitude of WGM frequency variations. It should be noted that the 95 kHz mode{'}s deformation gradient is an antisymmetric function about $Z=0$, and hence the mean strain is zero, compared to the symmetric 127 kHz mode. This results in effectively zero strain contribution to $\frac{df}{dx}$ in equation (\ref{eq:dfdx}) for the 95 kHz mode, hence the main component of transductance for this mode is expected to be a result from the physical change of boundary conditions.\\
\indent Following the treatment used by Tobar and Mann \cite{mann,krupka}, a mode matching technique is used to determine the $p_i$ values and the E-field distributions in each of the cylindrical coordinate directions. This can be done for both WGH and WGE modes with any number of azimuthal maxima. Then, taking the deformation gradients for the two mechanical modes produced in FEM, $M_i$ values are determined to finally produce an estimate for $\frac{df}{dx}$ as shown in Table \ref{tab:1}.

\begin{table}[h!]
	\begin{tabular}{c||c|c|c|c}
		\hline
		$\Omega_0/2\pi$ (kHz) & $M_z$ & $M_r$ & $M_\phi$ & $|df/dz|$ (MHz/$\mu$m) \\
		\hline
		94.97 & 0* & 0* & 0* & 0.085 \\
		127.07 & 0.38 & 0.0017 & 0.024 & 0.18\\
		\hline
	\end{tabular}
	\caption{Calculated values for the WGH$_{15,1,1}$ mode and the 95 kHz and 127 kHz mechanical modes. * a value of zero is due to the asymmetry of the 95 kHz mode strain curves.}
	\label{tab:1}
\end{table}

\section{Exciting Mechanical Modes and Measuring Q-Factors}

\begin{figure}[b!]
\centering
	\includegraphics[width=0.5\textwidth]{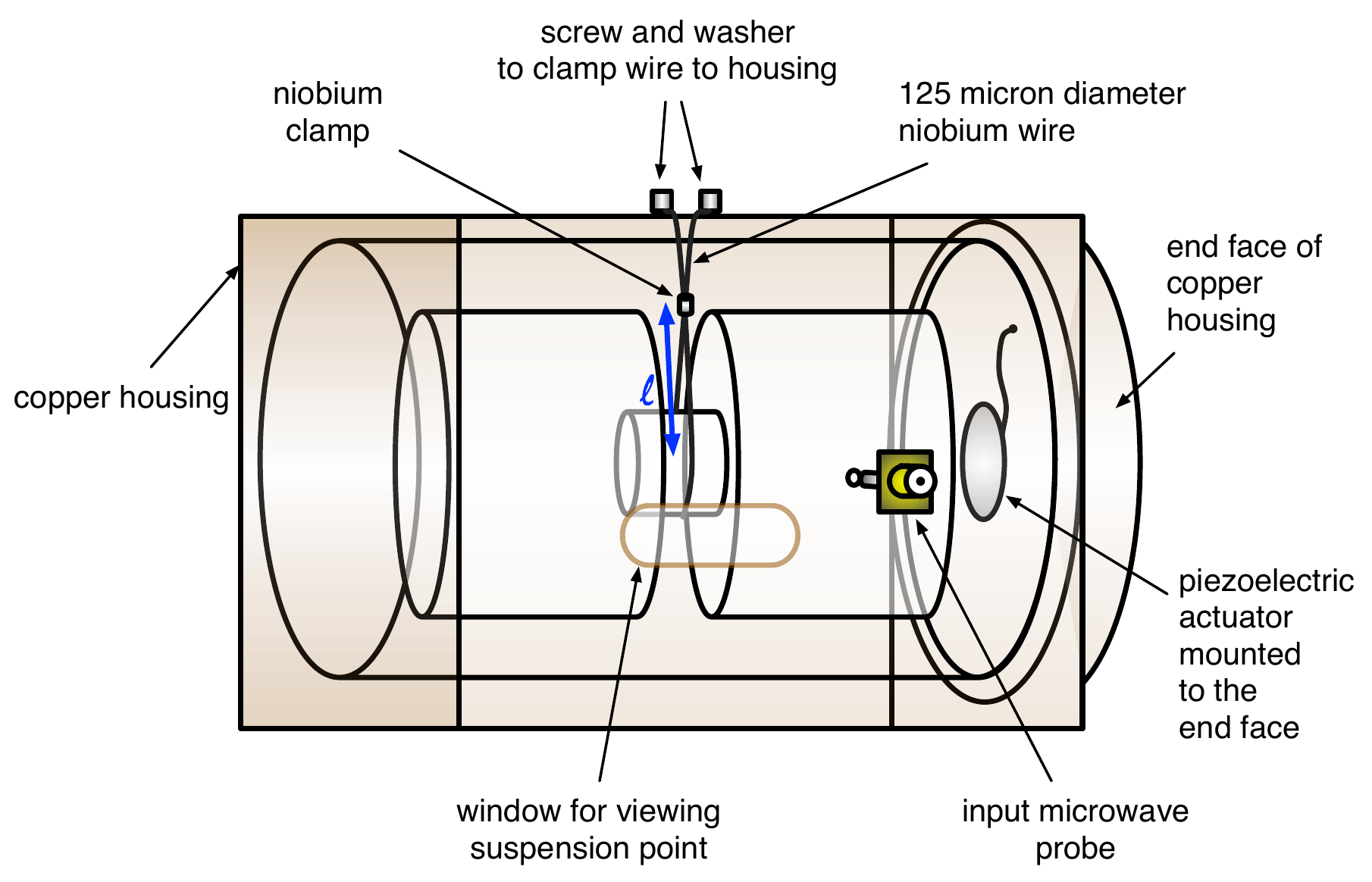}
	\caption{\small{A diagram of the SB suspended inside a copper cavity. The length of wire from the point of contact with the crystal neck to the position of the niobium clamp, labelled in blue, is the length that must be tuned.}}
	\label{fig:inside}
\end{figure}

\begin{figure}[t!]
\centering
	\includegraphics[width=0.45\textwidth]{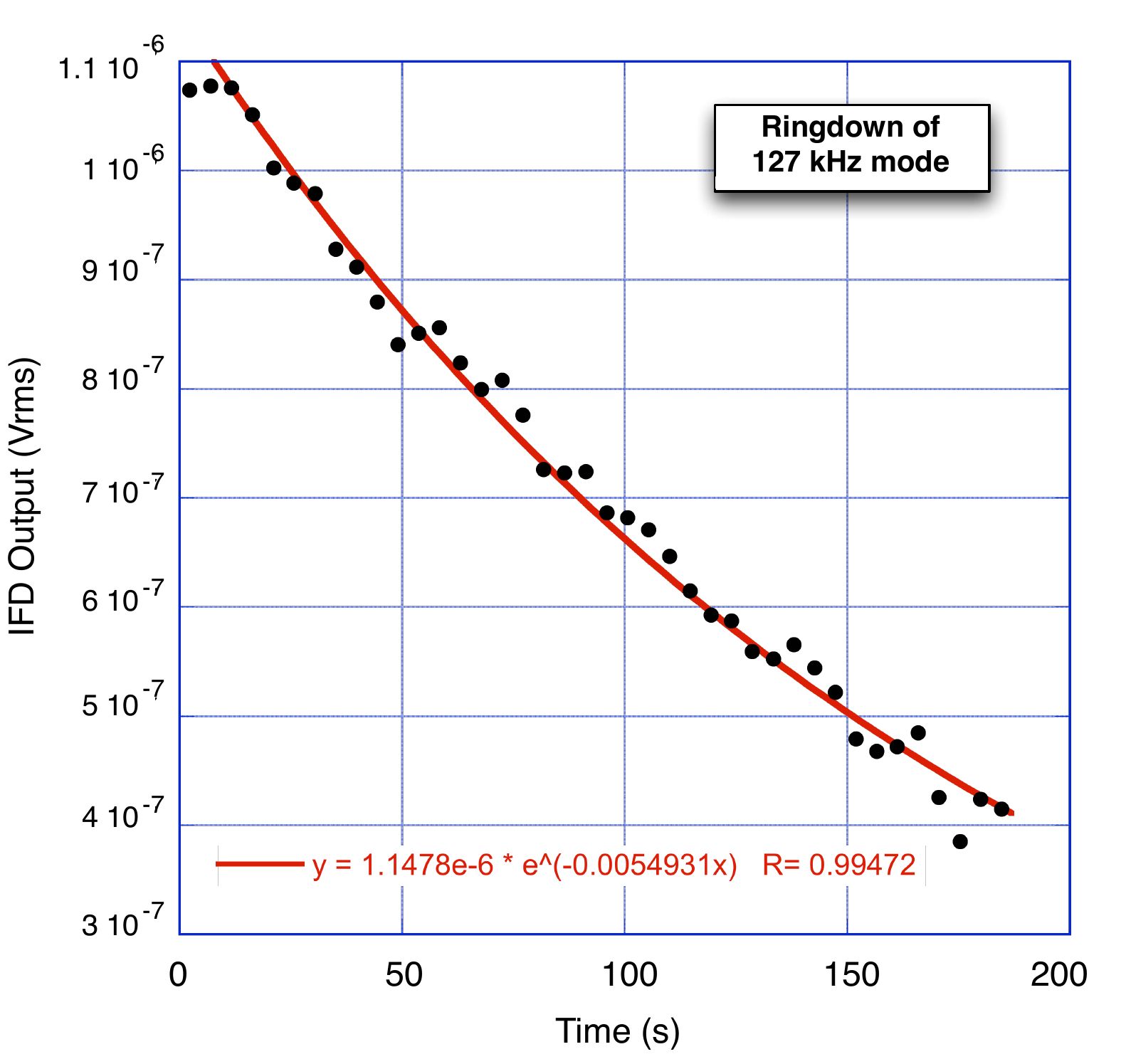}
	\caption{Mechanical Q-factor can be obtained from fitting the ringdown of the mechanical mode, measured by the output of the frequency discriminator.}
	\label{fig:ring}
\end{figure}

\noindent Initially, a piezoelectric actuator was used to excite the SB and therefore locate the resonant mechanical modes. The piezo was mounted to the SB's copper housing on a lid at one of the end faces, and is not in direct contact with the crystal. This setup is shown in Fig. \ref{fig:inside}. The piezo driving frequency was chirped over a small frequency range (on the order of 1 kHz) to conduct a relatively broadband search for resonances around the predicted frequencies. However, under vacuum the only path for the acoustic excitation of this form is through the suspension wire and if the length of this wire is optimally tuned such that the driving frequency is non-resonant with the wire, the transmission of energy to the bar will be minimal and the modes difficult to excite. In addition to this, driving the piezo with a periodic chirp results in the reduced contrast of the resonant peak viewed on a network analyser due to non-resonant excitation caused by shaking of the copper housing.\\
\indent An alternative method for mechanical excitation is applying radiation pressure force (RPF) to the crystal by modulating the power of the AM microwave signal at the mechanical resonance frequency (see Fig. \ref{fig:oscillator}). Once the AM is switched off the SB will continue to resonate mechanically with its amplitude ringing down with a characteristic time constant, $\tau$, which can be used to calculate the Q-factor of the mode ($Q_m=\pi \tau f_{mech}$). The results of this ringdown technique are shown in Fig. \ref{fig:ring}.\\
\indent When used together, these two techniques provide an effective and simple method for the excitation of mechanical modes and Q-factor measurements {--} the piezo is chirped to find the exact location of the resonance (within one bandwidth, on the order of mHz), and the input microwave signal is then AM{-}modulated at this resonant frequency for a period of time longer than the ring down time of the mode (to allow full excitation). The AM{-}modulation is then switched off, and finally the amplitude of the resulting peak is tracked over a time period of about 200 s. Fig. \ref{fig:ring} shows a measured value of $\tau=1/0.0052$ $s$ for the 127 kHz mode, corresponding to a mechanical Q-factor of  $7.7\times 10^7$.\\

\noindent By varying the amount of modulation on the incident power and measuring the peak amplitude immediately after AM{-}modulation has been switched off (provided RPF has been applied for a sufficiently long period of time that the resonator is in a quasi-static regime), one can experimentally determine a value for $df/dx$ for each of the mechanical modes, and compare with predicted values, as will be shown.\\

\indent Treating the mechanical sapphire resonator as a standard harmonic oscillator we have:

\begin{equation}
m_{\text{eff}}\frac{d^2x(t)}{dt^2}+m_{\text{eff}}\Gamma_m\frac{dx(t)}{dt}+m_{\text{eff}}\Omega_m^2x(t)=F_{ext}(t)\hspace{0.01\textwidth},
\end{equation}

which has the standard solution for displacement:

\begin{equation}
\delta x(\omega)=\frac{1}{m_{\text{eff}}}\frac{F_{ext}(\omega)}{\Omega_m^2-\omega^2-i\omega\Gamma_m}\hspace{0.01\textwidth}.
\end{equation}

\noindent Here, $m_{\text{eff}}$ is the effective mass of the bar in its resonant mechanical mode of frequency $\Omega_m$, $\Gamma_m=\Omega_m/Q_m$ is the mechanical decay rate, and $F_{\text{ext}}$ is some applied force. We can relate this displacement to a physically observable quantity, i.e. the voltage produced at the mixer output of the microwave readout system:

\begin{equation}
\delta u(\omega)=\delta x(\omega)\left(\frac{du}{df}\right)\left(\frac{df}{dx}\right)\hspace{0.01\textwidth}.
\end{equation}

\noindent We wish to solve this equation for the case of $F_{ext}$ originating from applied RPF. The expression for this can be found by differentiating the standard interaction Hamiltonian for an optomechanical system with respect to position \cite{optomech}:

\begin{equation}
\hat{H}_{int}=-\hbar g_0\hat{a}^\dagger\hat{a}\left(\hat{b}+\hat{b}^\dagger\right)\hspace{0.01\textwidth},
\end{equation}
 
where $\hat{a}^\dagger$ and $\hat{a}$ are the photon raising and lowering operators respectively, $\hat{b}^\dagger$ and $\hat{b}$ are the phonon raising and lowering operators respectively, and $g_0$ is the single photon optomechanical coupling:
	\begin{equation}
	\left. \begin{array}{lr}
\displaystyle g_0=G x_{\text{ZPF}}=\frac{d\omega}{dx}x_{\text{ZPF}}, \hspace{0.04\textwidth} & \displaystyle x_{\text{ZPF}}=\sqrt{\frac{\hbar}{2m_{\text{eff}}\Omega_m}} \hspace{0.01\textwidth} .
\end{array} \right. 
\label{ham}
	\end{equation}

\begin{figure}[t!]
\centering
\includegraphics[width=0.45\textwidth]{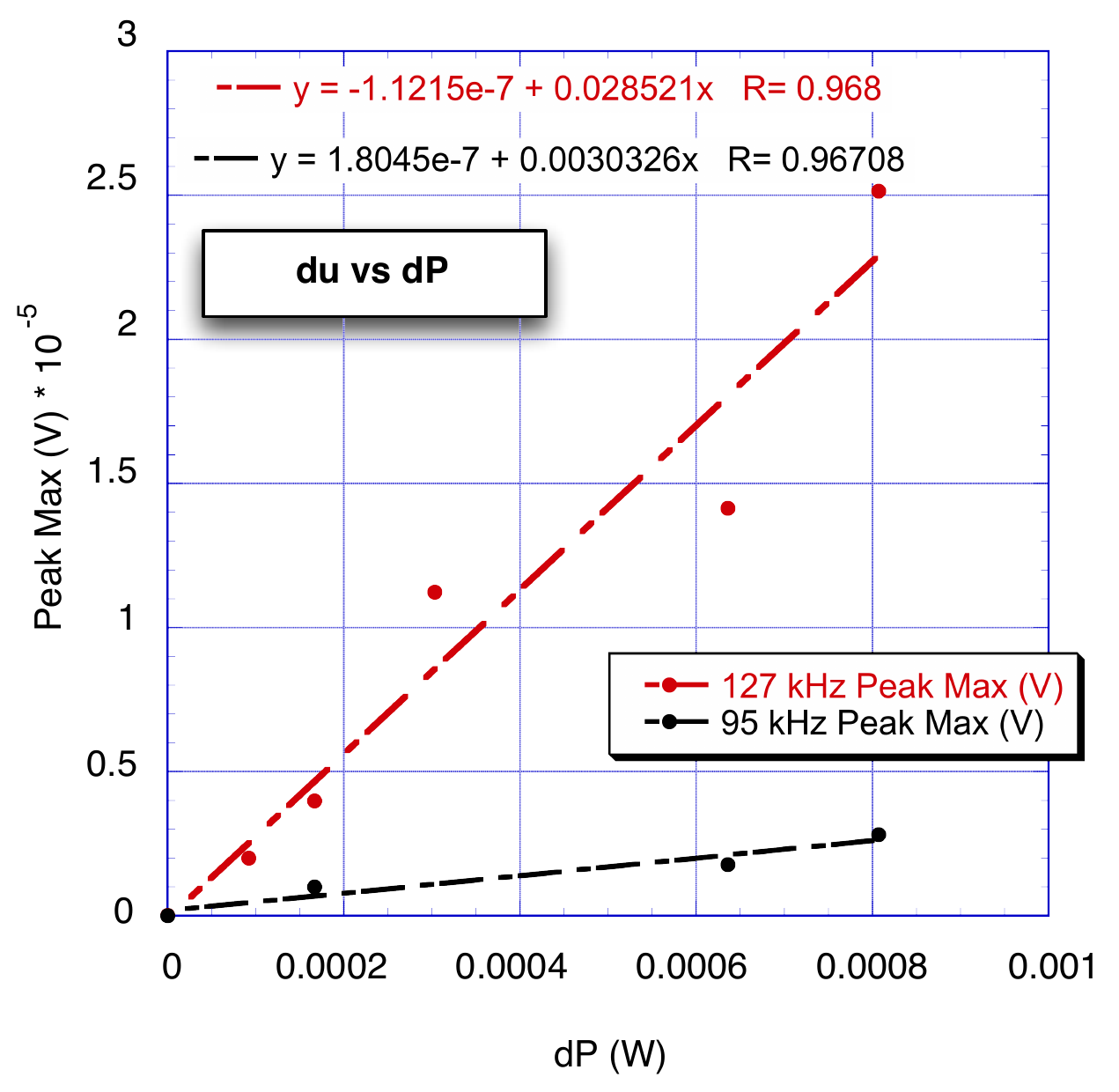}
\caption{Measurements of maximum peak amplitude of readout system output (proportional to displacement) to varying amounts of RPF.}
\label{fig:dudp}
\end{figure}

\noindent We can define the position operator $\hat{x}$ as 

\begin{equation}
\hat{x}=x_{\text{ZPF}}\left(\hat{b}+\hat{b}^\dagger\right)\hspace{0.01\textwidth},
\end{equation}

which will allow us to solve for $F_{ext}$ :

\begin{equation}
F_{ext}(t)=-\frac{d\hat{H}_{int}}{d\hat{x}}=\hbar G \hat{a}^{\dagger}\hat{a}=\hbar\frac{d\omega}{dx}n_{cav}\hspace{0.01\textwidth}.
\end{equation}
\noindent Here, $n_{cav}$ is the number of photons inside the cavity, and can be solved by dividing the total energy inside the cavity\cite{hartnett} by the energy of a single photon, as follows:

\begin{equation}
\begin{array}{c}
\displaystyle n_{cav}=\frac{E_{\text{WGM}}}{E_{\text{singlephoton}}}\hspace{0.01\textwidth}, \vspace{0.01\textwidth}\\
\displaystyle E_{\text{singlephoton}}=\hbar\omega\hspace{0.01\textwidth},\vspace{0.01\textwidth}\\
 \displaystyle E_{\text{WGM}}=P_{inc}\frac{Q_e}{\omega}\frac{4 \beta_1}{\left(1+\beta_1+\beta_2\right)^2}\frac{1}{1+4Q_e^2\left(\frac{\omega-\omega_e}{\omega_e}\right)^2}\hspace{0.01\textwidth},
\end{array}
\end{equation}

where $P_{inc}$ is the power of the incident microwave signal at frequency $\omega_e$, $Q_e$ the quality factor of the WGM, and $\beta_1$ and $\beta_2$ are the couplings between the WGM resonance and the input and output microwave probes, respectively. \\
\indent Assuming we are driving the microwave input at the WGM resonance ($\omega=\omega_e$), we finally arrive at an expression for the radiation pressure force inside the sapphire:
\begin{equation}
F_{ext}(t)|_{\omega=\omega_e}=\frac{d\omega}{dx}\frac{P_{inc}(t)Q_e}{\omega_e^2}\frac{4\beta_1}{\left(1+\beta_1+\beta_2\right)^2}\hspace{0.01\textwidth}.
\label{eq:f}
\end{equation}

\indent As a side note, this equation is identical to that obtained by Locke \textit{et al.}\cite{clayton2} when modelling the optomechanical system as an LCR circuit with a modulated capacitance. 

Assuming the incident power is modulated at $\Omega_m$, we can derive an equation for the voltage output by the microwave readout system:

\begin{equation}
\delta u (\Omega_m)=\pi\sqrt{2\pi}\left(\frac{du}{df}\right)\left(\frac{df}{dx}\right)^2\chi_{\text{WGM}}\chi_{mech}\delta P\hspace{0.01\textwidth},
\label{eq:final}
\end{equation}
where $\chi_{\text{WGM}}=\frac{4 \beta_1}{\left(1+\beta_1+\beta_2\right)^2}\frac{Q_e}{\omega_e^2}$ and $\chi_{mech}=\frac{Q_m}{m_{\text{eff}}\Omega_m^2}$.

From \hbox{Eq. \ref{eq:final}}, we can observe that by measuring $du/dP$, one can determine a value for $df/dx$ experimentally. The results of this measurement are shown in Fig. \ref{fig:dudp}, and every other value in Eq. \ref{eq:final} can be directly measured from the system. From the gradient of these two curves, $df/dx_{127}=0.19$ MHz/$\mu$m and $df/dx_{95}=0.05$ MHz/$\mu$m. \\
\indent The value for the 127 kHz mode is in very good agreement with the predicted value ($\sim 6\%$ error), while the 95 kHz mode is measured to be 40$\%$ smaller than its estimated value (see Table \ref{tab:1}). This can be attributed to the assumption that there was no strain induced contribution to the 95 kHz transduction in the simulation. As the WGM used has $\sim 95\%$ of its E-field in the z-direction, any strain contribution would arise mainly in this axis, which would subtract from the value obtained from purely dimensional changes (see Eq. \ref{eq:dfdx}). For example, a strain contribution 1/10$^{\text{th}}$ the magnitude of the 127 kHz mode would result in the predicted value being in good agreement with the measured value. This arises in practice because the strain curve of the 95 kHz mode is not a pure odd function, and the WGM is not a pure even function about $Z=0$ (see Fig. \ref{fig:Mz}), as was assumed when treating the interaction of the two modes with Eq. \ref{eq:dfdx}, which, in addition, is only an approximation to first order. Some small perturbations of these two symmetries will result in a small strain contribution to $df/dx_{95}$, which will subtract from the purely dimensional contribution.

\section{Optimising Mechanical Q-Factors}
\begin{figure}[h!]
\centering
\includegraphics[width=0.41\textwidth]{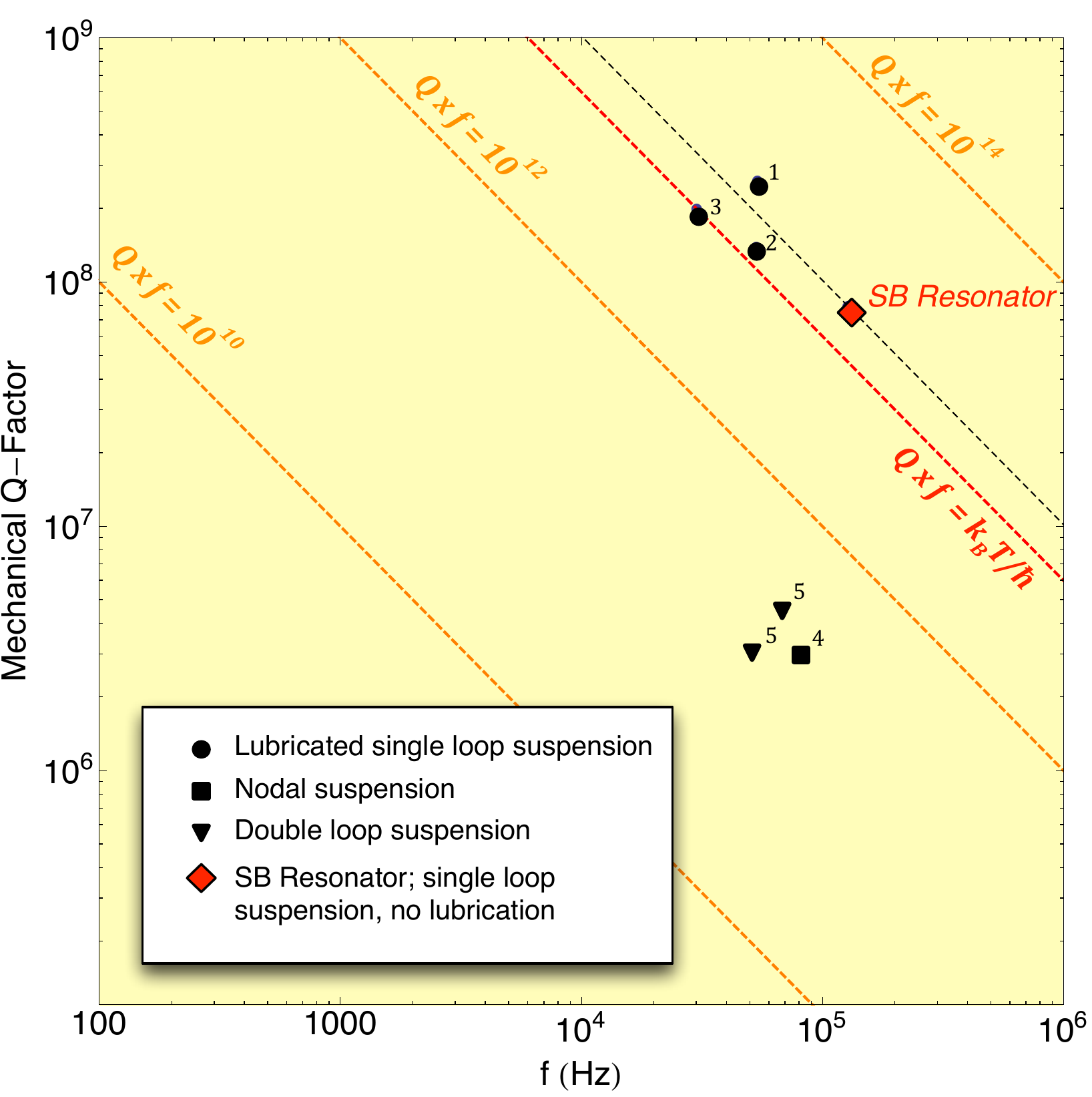}
\caption{Reported mechanical quality factors and resonant frequencies of published experiments with sapphire bar optomechanical systems at room temperature: 1. (Rowan \textit{et al.}, 2000)\cite{rowan}, 2. (Locke \textit{et al.}, 2001)\cite{claytonthesis}, 3. (Braginsky \textit{et al.}, 1985)\cite{braginsky}, 4. (Numata \textit{et al.}, 2000)\cite{numata}, 5. (Uchiyama \textit{et al.}, 1999)\cite{uchiyama}. The experiments are differentiated by their suspension techniques. The black dashed line represents the $Q\times f$ product of the SB.}
\label{fig:qf}
\end{figure}

\noindent Energy dissipated through the suspension is generally the dominant loss mechanism in optomechanical systems, as long as they are held under vacuum.\\
\indent The wire-loop suspension used to hold the SB is a method traditionally used by the gravitational wave community, which has been shown to achieve the highest mechanical Q-factors over other suspension schemes\cite{braginsky} (see the comparison of suspension techniques depicted in Fig. \ref{fig:qf}). Braginsky \textit{et al.} \cite{braginsky} modelled the losses in such a suspension as;

\begin{figure}[t!]
\centering
	\includegraphics[width=0.45\textwidth]{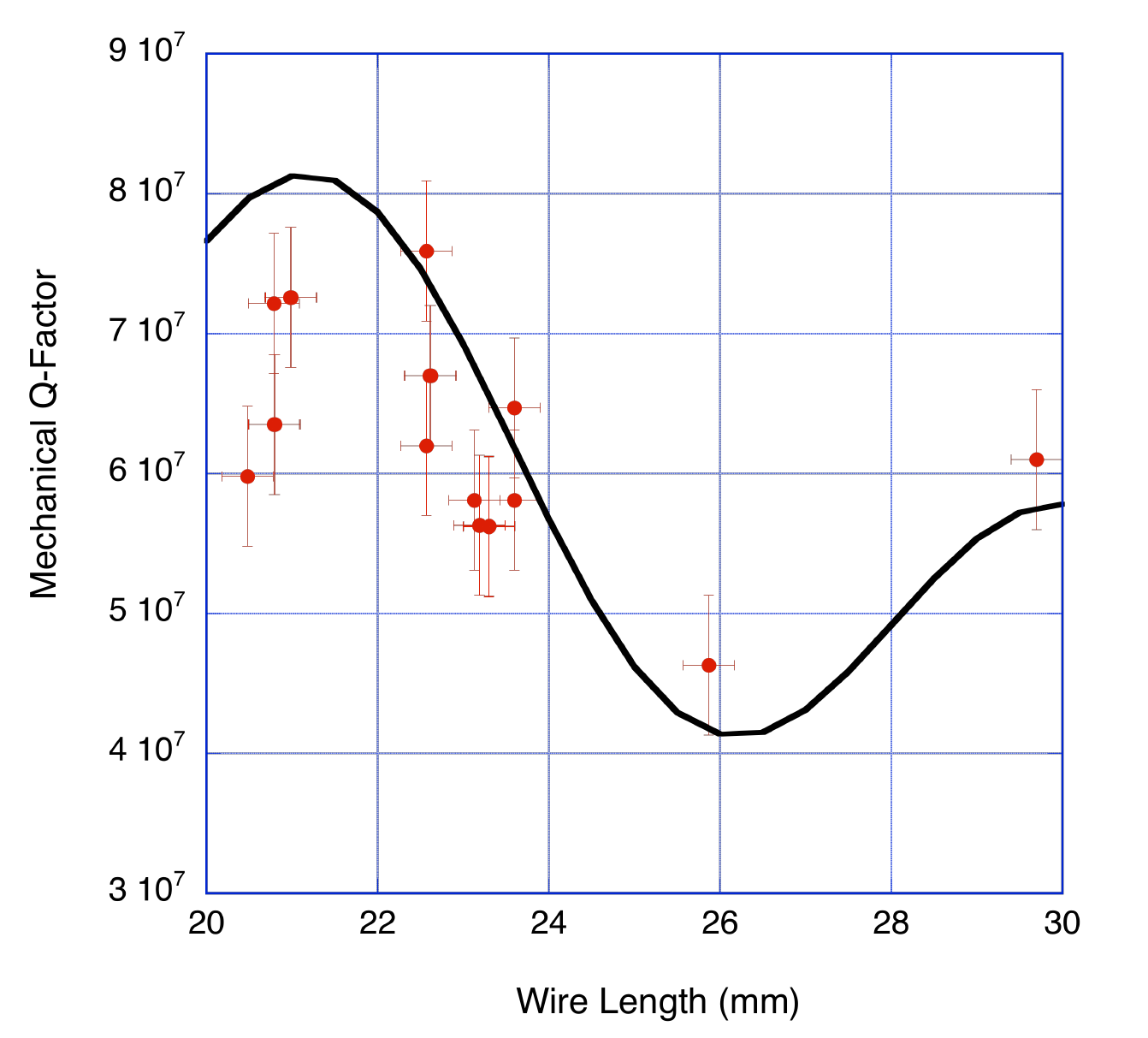}
	\caption{Measured mechanical Q-factors as length of Nb suspension wire is altered. Lengths are measured from the angle made by the suspension. A speed of sound in Nb of 2200 m/s produces the best fits, in agreement with quoted values for the speed of transverse waves in a thin Nb rod at 20 $^\circ$C.}
	\label{fig:qvsl}
\end{figure}

\begin{multline}
\frac{1}{Q_{s}}=\frac{2\rho_w A_w l_w}{m_{\text{eff}} Q_w}\frac{x_{ext}^2}{x_0^2}\times \\
\left[1+\frac{1}{2}\left(\frac{\omega_m l_w}{v_w Q_w}\right)^2-\cos\left(\frac{2l_w\omega_m}{v_w}\right)\right]^{-1}
\label{eq:susp}
\end{multline}

\noindent where $\rho_w$, $A_w$, $l_w$, $Q_w$ and $v_w$ are the density, cross-sectional area, length, mechanical Q-factor and velocity of sound of the wire, respectively, and $x_0$ and $x_{ext}$ are the amplitude of vibration of the resonator and the amplitude at the point of contact with the wire, respectively. Equation \ref{eq:susp} predicts a periodic variation in Q-factors as the length of the wire (shown in blue in Fig. \ref{fig:inside}) is changed. However, Eq. \ref{eq:susp} has only been shown to provide qualitative agreement with experimental results, as the loss mechanism is far more complicated than this simple model. \\

\begin{figure}[t!]
\centering
		\includegraphics[width=0.45\textwidth]{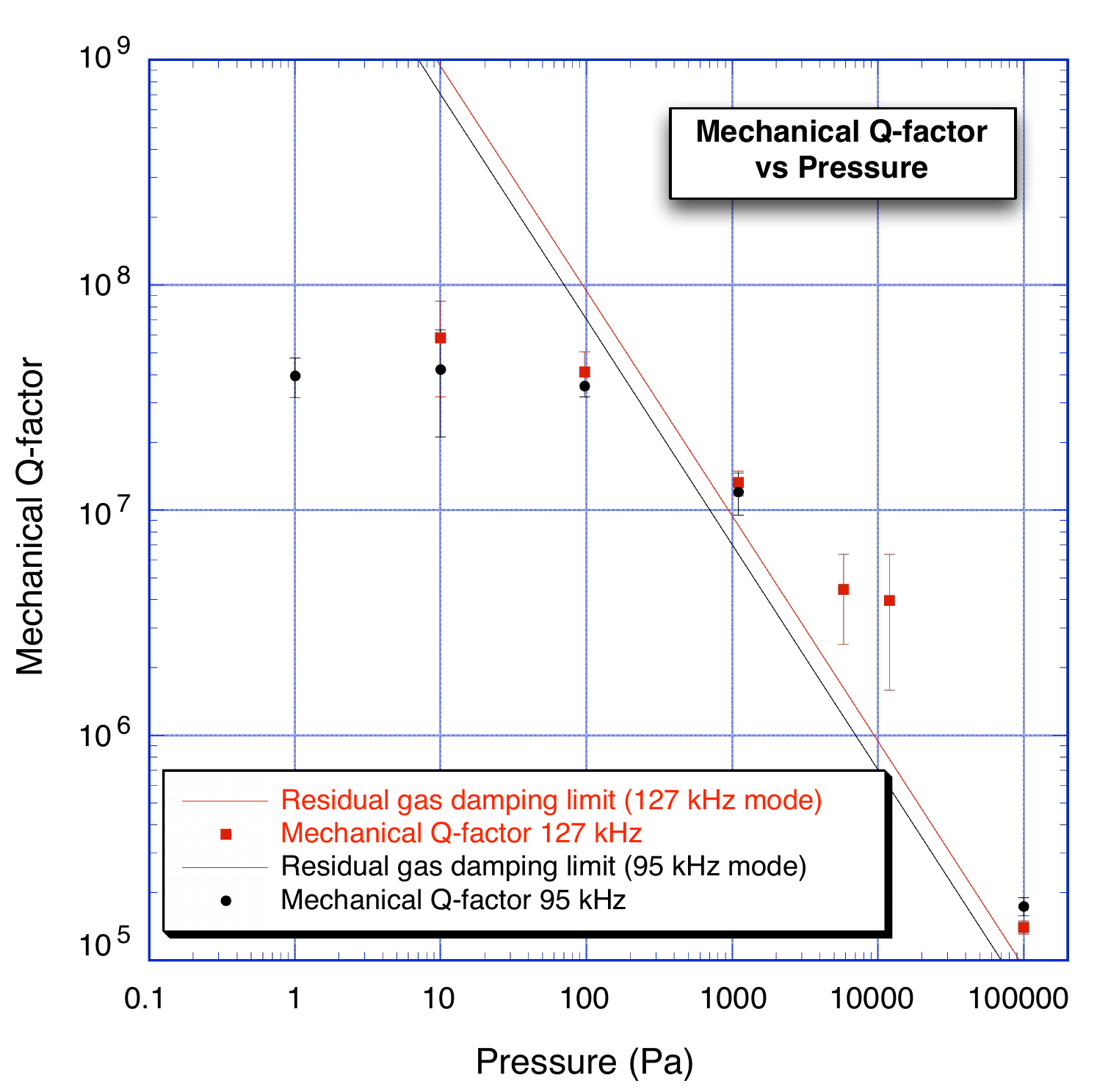}
		\caption{Mechanical Q-factor of the 127 kHz and 95 kHz modes vs pressure. Residual gas damping limits as predicted by equation (\ref{eq:gas}).}
		\label{fig:pressure}
\end{figure}

\noindent Two different wires were trialled for the suspension system; an 80 $\mu$m diameter tungsten wire and a 125 $\mu$m niobium wire, with better results being achieved with the latter due to its higher intrinsic $Q_w$. The niobium clamp (shown in Fig. \ref{fig:inside}) is necessary to keep enough friction around the neck of the SB to maintain balance, but can also be used to alter the length of the wire by moving its position. The clamps are made by cutting small lengths of 0.65 mm diameter niobium tubing. Tests were made with both clamped and unclamped pieces of tubing, with the former producing higher Q-factors due to the boundary condition in the wire resonance being fixed.\\
\indent Results of varying wire length are presented in Fig. \ref{fig:qvsl}. Each experimental data point represents an entirely new wire suspension, as the clamping technique used fixes the wire{'}s length. As such, it is possible for surface conditions and the loop's position along the z-axis to change between measurements. Therefore each of the data points in Fig. \ref{fig:qvsl} represents a lower limit of Q-factor for any given wire length.\\
\indent 
Braginsky \textit{et al.} \cite{braginsky} states that there is only a qualitative agreement between Eq. \ref{eq:susp} and the experimental data, with measured Q-factors being larger than the predicted values by almost a full order of magnitude. This is mainly due to resonator vibrations at the point of contact with the suspension wire being only partially transmitted to the wire. As such, we insert an additional term inside the square brackets of Eq. \ref{eq:susp} to represent coupling between the wire and crystal. It takes the form of $1/\xi$. It should also be noted that the second term inside the square brackets of Eq. \ref{eq:susp} is much less than 1 and can therefore be neglected. The experimental results are then fitted with the following equation:

\begin{equation}
Q_{s}=\frac{m_{\text{eff}} Q_w}{2\rho_w A_w l_w}\frac{x_0^2}{x_{ext}^2}\left[1+\frac{1}{\xi}-\cos\left(\frac{2l_w\omega_m}{v_w}\right)\right],
\label{eq:susp2}
\end{equation}

\noindent which is shown in black in Fig. \ref{fig:qvsl} and provides a good bound of the measured Q-factors given the fitted values of $Q_w=8.5\times10^5$ and $\xi=0.31$.\\
\\
\noindent Fig. \ref{fig:pressure} shows the dependence of mechanical Q-factors on pressure, as predicted by the following equation: \cite{gas}
\begin{equation}
\frac{1}{Q_{gas}}\approx\frac{PA}{M \omega_0}\sqrt{\frac{\mu}{k_bT}},
\label{eq:gas}
\end{equation}
where P is pressure (in Pascals), A the surface area of the mechanical resonator, M its mass and $\mu$ is the mass of molecule comprising majority of the gas in question.
As can be seen from Fig. \ref{fig:pressure}, below a certain threshold pressure, residual gas damping is no longer the dominant loss mechanism, with Q-factors of close to $10^8$ demonstrated. \\

Throughout the literature regarding wire loop suspensions, it is often noted that by applying a small layer of lubrication (most commonly animal fat is used) to the surface of the mechanical resonator at the point of suspension, significant improvements in Q-factors can be achieved \cite{braginsky,claytonthesis,rowan}. Previously, at room temperature, Q-factors on the order of $10^8$ and $Q_m\times f \ge 6\times 10^{12}$  have only been achieved with some form of lubrication, and this is therefore a strategy that could be implemented in the immediate future of this work. However even without lubrication, the present work corresponds to a $Q_m\times f$ product of $10^{13}$, equivalent to the best measured values of sapphire acoustic systems, as depicted in Fig. \ref{fig:qf}. This product is considered a figure of merit for acoustic resonant systems, as it is a direct measure of the degree of decoupling from the thermal environment. Specifically, $Q_m \times f > k_b T/h$ is the condition for neglecting thermal decoherence over one mechanical period, and is considered the minimum requirement for quantum optomechanics \cite{optomech}. At room temperature, this lower limit is equal to $6\times10^{12}$, and is represented in Fig. \ref{fig:qf} by the red dashed line. 

Importantly, our results are consistent with the best reported values measured at room temperature in sapphire, most likely limited by the quality of the crystal \cite{braginsky,claytonthesis}. Thus, this result demonstrates that equivalent performance to state-of-the-art sapphire acoustic systems can be achieved with a modified cylindrical structure. All other results presented in Fig. \ref{fig:qf} are achieved using modes in right cylinders as shown in the right-hand side limit of Fig. \ref{fig:mechanical}. This is the first time optomechanics in a sapphire bar has been attempted with a modified architecture.\\

\indent By cooling the SB, the coupling between the suspension and crystal will decrease, resulting in the suspension losses decreasing. Q-factors on the order of $10^9$ have been achieved using wire suspended sapphire at liquid helium temperatures\cite{braginsky}. So whilst a maximum value of $\sim8\times10^7$ has been achieved thus far, we are optimistic of improving this.

\section{Microwave Readout System}
\label{sec:ro}

\begin{figure*}[t!]
	\centering
		\includegraphics[width=0.8\textwidth]{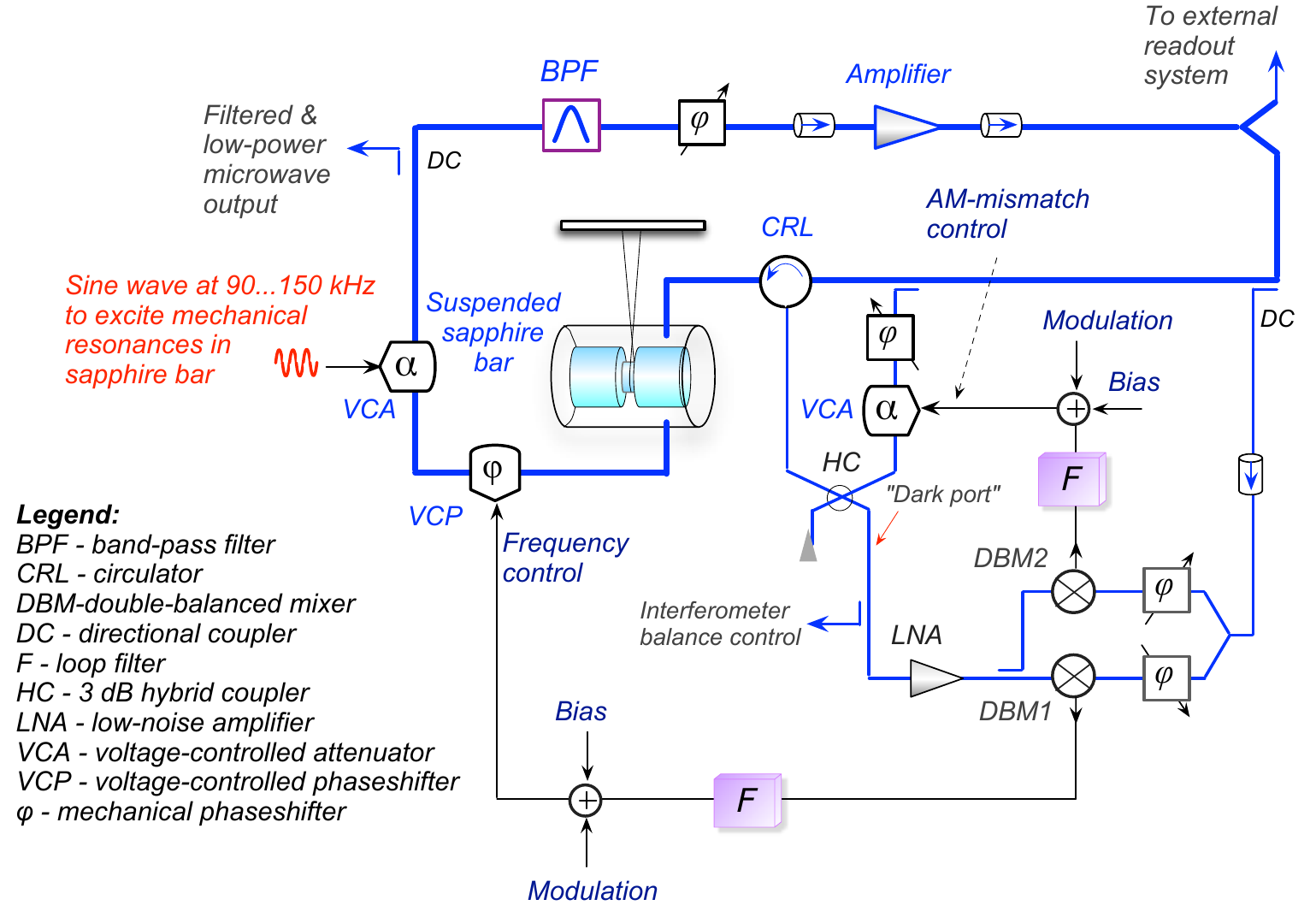}
		\caption{Schematic diagram of a frequency stabilized microwave oscillator based on suspended sapphire SB resonator excited in electromagnetic modes of whispering gallery.}
		\label{fig:oscillator}
		\hrule
\end{figure*}

\subsection{Detection of Frequency Fluctuations}
As described in the previous sections, a freely suspended sapphire bar is a unique physical object capable of supporting very high quality resonances, both mechanical and electromagnetic. It offers an opportunity to conduct a detailed study of the bar{'}s mechanical resonances via their influence on the microwave resonances. In addition, one can investigate the inverse effects of the microwave readout on the SB{'}s mechanical properties including the degeneration (cold damping) and regeneration (parametric excitation) of elastic vibrations \cite{braginsky,veitch2}.

This section describes a microwave readout system for monitoring the vibration state of a freely suspended sapphire bar. We expected that the sensitivity of the microwave readout would be sufficiently high to permit the very first observations of the bar's mechanical resonances excited by thermal fluctuations of the sapphire crystalline lattice. 

There are at least two techniques, which can be employed for the microwave assisted detection of elastic vibrations of the SB. Both techniques are widely used in the field of oscillator frequency stabilisation for high-resolution measurements of fast frequency fluctuations\cite{ivanov,ivanov2}. In one case, the SB can be configured as a dispersive element of a microwave frequency discriminator. When driven from a fixed frequency signal source, the SB acts as a frequency-to-voltage converter producing voltage varying synchronously with resonant frequency of given WG-mode excited by the source. As a result, the task of analysing spectrum of a microwave signal is reduced to computing the Fast Fourier Transform of a sampled voltage at the discriminator output.
Alternatively, the SB can serve as a band-pass filter of a self-exciting microwave loop oscillator \cite{leeson}. This would {``}imprint{''} the spectrum of the SB mechanical normal modes on to the spectrum of the microwave signal. Once again, a frequency discriminator can be used to convert frequency fluctuations of the microwave signal into synchronous fluctuations of voltage.
Performance-wise, both of the above-mentioned techniques are identical; in each case the useful frequency fluctuations associated with elastic vibrations of the SB must compete with the same spurious fluctuations of the microwave readout electronics; and each technique, given some optimisation, could potentially lead to spectral resolution close to the Standard Thermal Noise Limit \cite{ivanov3}.

\subsection{Microwave Sapphire Bar Oscillator}

Figure \ref{fig:oscillator} shows a schematic diagram of a microwave loop oscillator based on the SB. The oscillator is frequency-locked to a given WGM of the SB to facilitate detection of the useful frequency fluctuations. The spectrum of the free-running oscillator at Fourier frequencies of interest ($\sim$ 100 kHz) is completely dominated by 1/$f$ phase noise of the microwave loop amplifier. These spurious fluctuations are suppressed by the frequency lock (or frequency control) system as explained below.

The key element of the oscillator frequency control system is an ultra-sensitive frequency discriminator. It consists of a microwave Mach-Zehnder interferometer with the suspended sapphire bar resonator and a phase sensitive readout system featuring low-noise amplifier (LNA) and a double-balanced mixer (DBM1 in Fig. \ref{fig:oscillator}). The microwave signal reflected from the SB interferes destructively with a fraction of the incident signal at the interferometer {``}dark port{''}. This cancels the carrier of the difference signal while preserving noise modulation sidebands resulting from 1/$f$-noise of the microwave loop amplifier. The residual noise at the dark port is amplified and demodulated to the baseband, producing an error voltage proportional to oscillator frequency fluctuations. The error voltage, after appropriate filtering, is applied to the electronic phase-shifter (VCP) in the microwave loop. This steers the oscillator frequency to that of the resonator, or more precisely, to the frequency at which the carrier was suppressed, which is typically well within the resonator bandwidth. 
The frequency discriminator, loop filter and VCP form the frequency control loop. If the control loop gain is sufficiently high, the fidelity with which oscillator frequency follows that of the resonator is determined only by technical fluctuations in the electronics of the frequency discriminator. In this respect, interferometric frequency discriminators are far superior to their conventional counterparts, as they exhibit effective noise temperature close to the ambient temperature, and are capable of handling much higher power levels\cite{ivanov3}. 

It is not difficult to understand why frequency discriminators with the highest sensitivity are based on spindle-shaped sapphire crystals fixed rigidly inside the protective metal shields. The main reason for this is the low vibration sensitivity of such resonators, which makes the task of carrier suppression fairly straightforward. The situation is different in the case of the SB whose rocking motion upsets both amplitude and phase balance of the interferometer. To cope with the SB rocking motion, the oscillator in Fig. \ref{fig:oscillator} features an additional feedback control system charged with the task of minimising amplitude mismatch between two signals interfering at the dark port. The error voltage for this feedback system is produced by the second mixer (DBM2) tuned in quadrature relative to the mixer of the frequency control loop. The voltage-controlled attenuator (VCA) in the interferometer arm completes the feedback loop acting as an actuator of the amplitude mismatch control system. 

The quadrature tuning of both DBMs is required to avoid the cross-talk between the frequency and amplitude control loops. Yet, the tolerances of such tuning proved to be not very stringent. For each control system, it was sufficient to roughly maximise the amplitude of the error signal (in response to some deliberately introduced perturbation) before closing the feedback loop.

An additional VCA in Fig. \ref{fig:oscillator} is used to modulate the amplitude of microwave signal. When the modulation frequency coincides with the frequency of mechanical resonance, parametric excitation of the normal mode ensues. This VCA was introduced when the refined suspension system made piezoelectric excitation of mechanical resonances ineffective. 

\begin{figure*}[t!]
	\centering
		\includegraphics[width=0.8\textwidth]{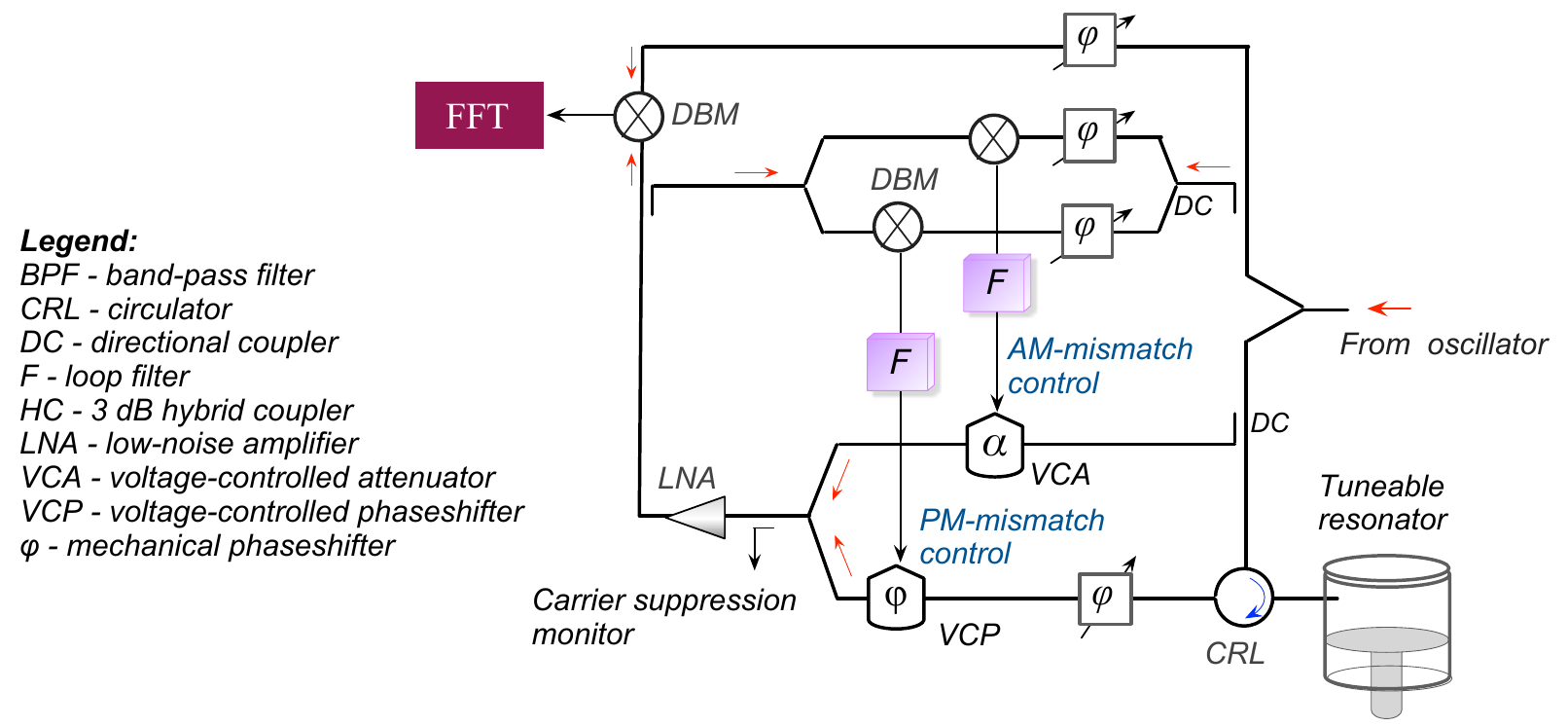}
		\caption{Schematic diagram of interferometric frequency discriminator with automatic carrier suppression.}
		\label{fig:IFD}
		\hrule
\end{figure*}

\subsection{Inteferometric Frequency Discriminator}

Approximately half of the power generated by the SB oscillator (Fig. \ref{fig:oscillator}) is diverted to the external frequency discriminator shown schematically in Fig. \ref{fig:IFD}. It is practically identical to the previously described built-in discriminator of the SB oscillator - it also contains a microwave interferometer with actively controlled balance to cope with frequency variations of the input signal induced by the rocking motion of the SB and changes of ambient temperature.

The dispersive element of the external frequency discriminator is a cylindrical metal cavity with a Q-factor of 15,000 and frequency tuning range of approximately 2 GHz. The wide tuning range of the cavity enables easy switching from one WGM of the sapphire bar to another, if one needs to investigate how a WGM of different polarisation and azimuthal number responds to vibration. 

\begin{figure}[t!]
	\centering
		\includegraphics[width=0.45\textwidth]{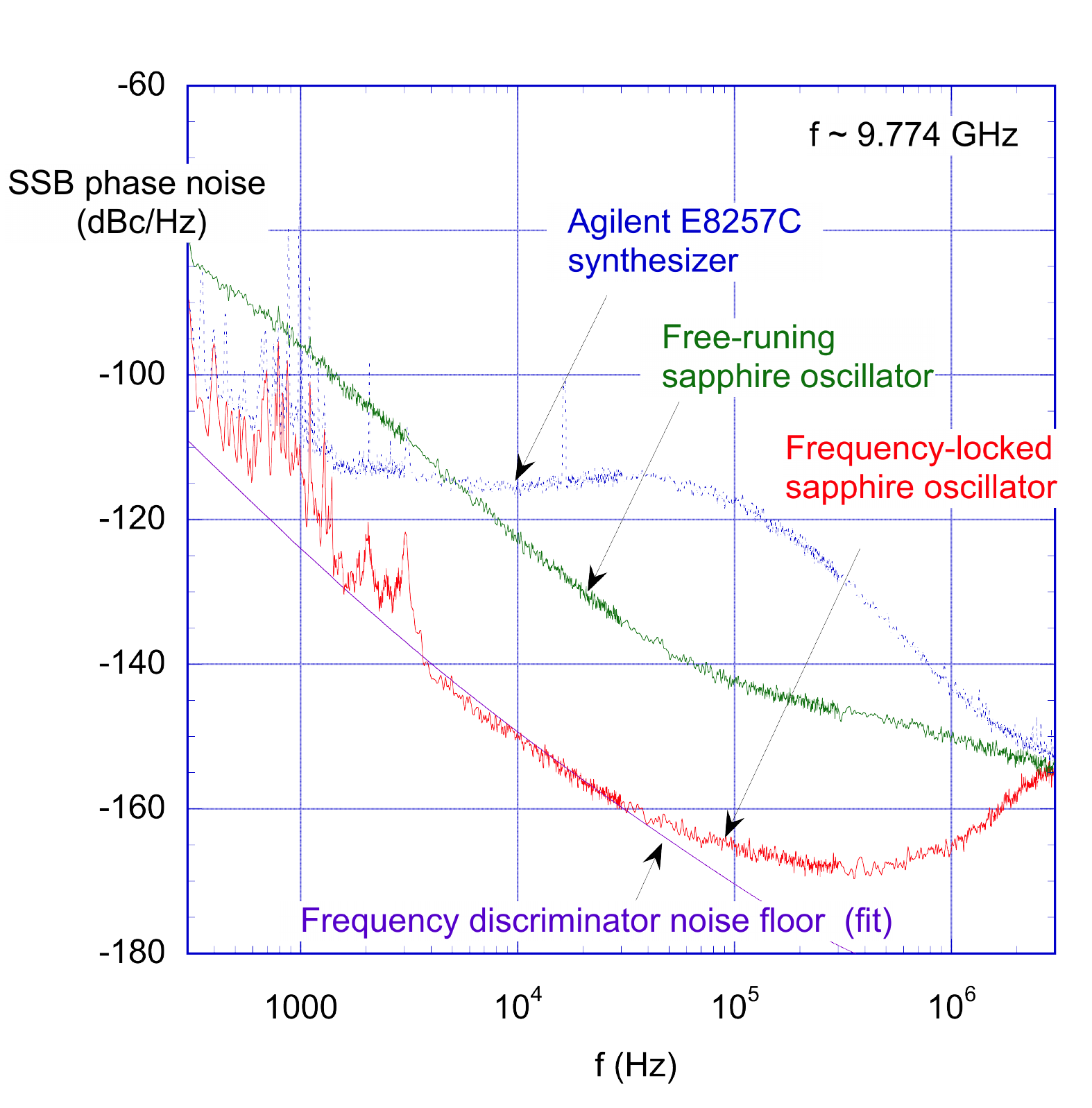}
		\caption{SSB phase noise spectra of various microwave signal sources measured with interferometric frequency discriminator with automatic carrier suppression. Power of the input signal is 19 dBm.}
		\label{fig:noise}
\end{figure}

Fig. \ref{fig:noise} shows spectra of phase fluctuations of various signal sources measured with the external frequency discriminator. The measurements were made at 9.774 GHz corresponding to the excitation of the WGH$_{15,1,1}$ mode of the SB. 

The power spectral density of phase fluctuations $S_\phi$ was inferred from that of the voltage noise $S_u$ via the following relationship:

\begin{equation}
S_\phi(F)=\frac{S_u(F)}{K_{FD}^2}\left(\frac{1}{\Delta f_L^2}+\frac{1}{F^2}\right)
\end{equation} 					

where F is the Fourier frequency, $\Delta f_L$ is the half-loaded bandwidth of the cavity resonator and $K_{FD}$ is the frequency discriminator {``}DC Ð sensitivity{''} measured at $F \ll \Delta f_L$. 

The bottom trace in Fig. \ref{fig:noise} shows the fit to the noise floor of the frequency discriminator expressed in the single sideband (SSB) units of dBc/Hz. The noise floor was measured with the cavity resonator replaced by a 50 $\Omega$ termination. Next, we measured phase noise of a commercial frequency synthesizer (Agilent E8257C). The idea was to verify that the voltage-to-phase conversion procedure we followed was correct (our results proved to be consistent with the specs of the Agilent instrument). Finally, we characterised the phase noise of the SB oscillator, both in the free-running and frequency-locked regimes. At Fourier frequencies 50 kHz $<$ F $<$ 500 kHz, we observed more than 20 dB of phase noise suppression when the frequency lock was engaged.

As follows from Fig. \ref{fig:noise}, the SSB power spectral density of spurious phase fluctuations is -165 dBc/Hz at F=100 kHz. The corresponding level of rms frequency fluctuations is 

\begin{equation}
\delta f=F\sqrt{S_\phi(F)}\sim 8 \times 10^{-4} Hz/\sqrt{Hz}
\end{equation}

Recalling the frequency-displacement sensitivity of the sapphire bar ($f_z \sim$ 0.18 MHz/$\mu$m) yields the displacement noise floor: $\delta z \sim \delta f/ f_z \sim 4 \times 10^{-15}$ m/$\sqrt{\text{Hz}}$. This corresponds to the highest displacement sensitivity ever reported in experiments with sapphire bar resonators as displacement transducers. Yet, the level achieved is still approximately twice as large as the 127 kHz mode{'}s elastic vibrations driven by thermal noise, hence Brownian motion is currently unobservable with the present configuration. 

At this stage, two questions can be posed: {``}What limits spectral resolution of the current experimental setup and how can it be improved?{''} In answering these questions, we, first of all, can single out two factors, which are almost equally responsible for the present level of displacement sensitivity. One of them is the relatively high level of phase noise of the frequency-locked microwave oscillator at F $\sim$ 100 kHz. This is because the current frequency control system starts loosing gain at Fourier frequencies above 10 kHz. Solving this problem will involve the design of a new low-pass filter of the frequency feedback loop based on operational amplifiers with gain-bandwidth product exceeding 1 GHz.

The second limitation arises from the comparatively low electrical Q-factor of the hollow metal cavity. We believe, that at least an order of magnitude improvement in displacement sensitivity can be gained by replacing the hollow metal cavity with a sapphire loaded cavity resonator. 

One {``}side effect{''} of resonator substitution is the loss of the wide frequency tunability. Yet, some residual tunability would remain owing to the relatively high sensitivity of sapphire resonators to temperature (df/dT $\sim$ 0.5{--}0.7 MHz/K depending on the type of WGM used). Another complication of resonator substitution is related to the narrow bandwidths of sapphire resonators. An additional control system would be required to keep the sapphire resonator {``}in sync{''} with the incoming signal. We plan to address this issue by controlling microwave power dissipated in the sapphire crystal as in \cite{ivanov4,ivanov5}, where such a technique was used to enable the phase referencing of a {``}slave{''} sapphire oscillator to the {``}master{''}. It should be remembered that the measurement sensitivity improves as a square root of the power of the SB oscillator.\\

\noindent A major advantage of the SB{'}s dumbbell-shaped architecture is that it is in fact two electromagnetic oscillators, which are both undergoing the same mechanical fluctuations. One can easily imagine a cross correlation scheme in which two readout systems as described above were constructed around either end of the SB. The primary mixer outputs of these readouts could then be cross correlated to eliminate uncorrelated electronic noise and boost the correlated mechanical signal, which would potentially allow never-before-seen levels of displacement sensitivity in such a system.

\section{Parametric Effects}
\label{sec:para}

\begin{figure}[b!]
	\centering
	\includegraphics[width=0.45\textwidth]{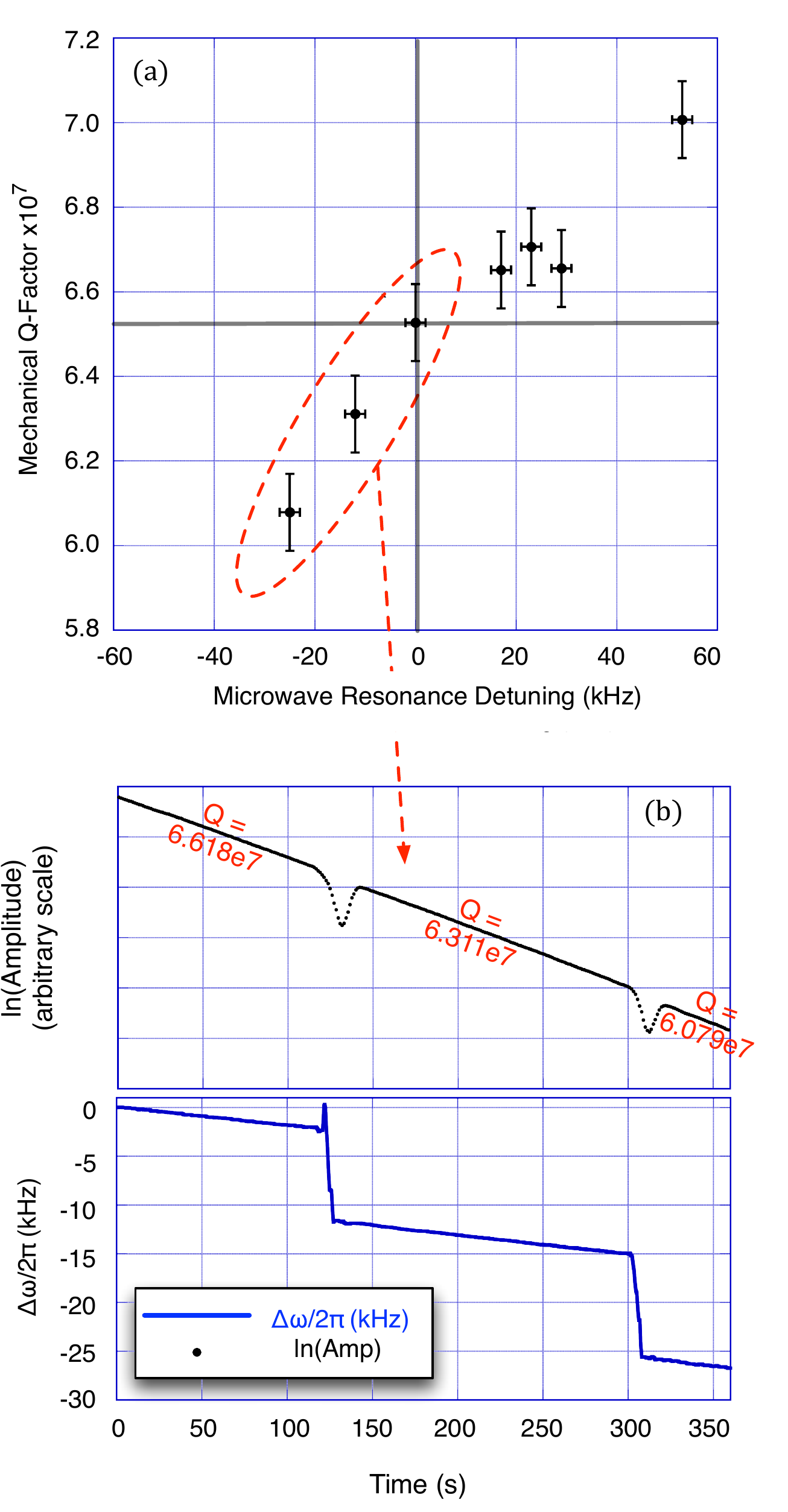}
	\caption{(a) Q-factor vs input microwave frequency detuning. The circled data points are obtained from analysing (b); a real time ringdown measurement of the 127 kHz mechanical mode as negative detuning is increased during the measurement. The natural log of the amplitude is taken such that the time constant $\tau$ will be equal to the inverse of the gradient of a linear fit for each detuning value{'}s data.}
	\label{fig:para}
\end{figure}

\noindent Parametric behaviour will always arise when a driven resonant system is coupled to a second resonant system. A phase difference between the mechanical oscillator and the microwave resonance will produce either a damped or driven system. As such, one can expect changes in the mechanical quality factor of a resonant optomechanical system as the {``}pump{''} signal is detuned from resonance. These effect have been well described in the past with various types of transducers \cite{crystal1,optomech,clayton2,braginsky2,tobarthesis}. The phenomenon, referred to as the {``}optical spring{''} effect, results from a Stokes/anti-Stokes process, whereby pumping above the optical resonance produces additional phonons; increasing the mechanical mode{'}s effective temperature and Q-factor, whilst pumping below resonance removes phonons from the system, effectively increasing losses and cooling the system\cite{teufel}. There should also be a corresponding mechanical frequency shift associated with microwave pump detuning, however it is predicted that it would be on the order of mHz, far below the resolution of the measurement technique used here, which relies upon a vector signal analyser{'}s {``}peak trace{''} function to record amplitude vs. time.\\
\\
\indent Figure \ref{fig:para}(a) demonstrates the {``}optical spring{''} effect for the 127 kHz mode of the SB resonator. These measurements were taken using a method depicted in Fig. \ref{fig:para}b, in which whilst the mechanical resonance rings down, the microwave detuning is changed from zero, and the resulting change in time constant is measured. Positive detuning results were also taken and the trials repeated. The transient response that can be seen immediately after a frequency shift in Fig. \ref{fig:para}b is a result of the readout system requiring a re-balance. \\
\indent Within the explored detuning range, $\Delta \omega_e < \Omega_m$, hence one expects an approximately linear relationship between Q-factor and detuning. The reason for this limited range is due to the restrictions placed by the bandwidth of the frequency control system acting on the loop oscillator (see Section \ref{sec:ro}). The loop oscillator and control system used to readout the mechanics of the SB complicate the microwave system beyond a simple LCR model. \\
\indent The presence of parametric effects is a promising result for the system, as it is only through resolved sideband cooling \cite{teufel} that a system such as the SB resonator could overcome thermal noise to reach a quantum limited state \cite{optomech,teufel}. \\
\\
\noindent To date, we have achieved excellent agreement between modelled and measured frequencies of a variety of mechanical modes of the SB resonator, and developed a robust method of mechanically exciting it via a piezoelectric shaker and via radiation pressure. The later method has also provided a method of calibrating the transducer{'}s frequency sensitivity to displacement, $df/dx$, which has been measured to good agreement with predicted results. The mechanical Q-factors have been optimised to a point at which they agree with maximum reported values at room temperature within limitations set by the quality of the sapphire, as does the $Q_m\times f$ product. Parametric back-action resulting in mechanical damping and excitation have also been observed. Finally, a novel microwave readout system has been constructed, which provides extremely low phase noise performance. This work provides an important enabling step for the next generation of kilogram scale mechanical oscillator experiments designed to measure and test the standard quantum limit and potentially investigate the nature of quantum gravity, thus proving that novel architectures of sapphire systems can still produce state-of-the-art results.\\

\end{document}